\documentclass[prc,amsmath,amsfonts,amssymb,twocolumn,floatfix]{revtex4-1}
\usepackage{graphicx}
\usepackage{bm}
\usepackage{color}
\usepackage[mathscr]{eucal}

\newcommand{\N}{\mathcal{N}}
\renewcommand{\P}{\mathcal{P}}

\begin{document}
\title{Quasiparticle Coupled Cluster Theory for Pairing Interactions}
\author{Thomas M. Henderson}
\affiliation{Department of Chemistry and Department of Physics and Astronomy, Rice University, Houston, TX 77005-1892}

\author{Jorge Dukelsky}
\affiliation{Instituto de Estructura de la Materia, CSIC, Serrano 123, E-28006 Madrid, Spain}

\author{Gustavo E. Scuseria}
\affiliation{Department of Chemistry and Department of Physics and Astronomy, Rice University, Houston, TX 77005-1892}

\author{Angelo Signoracci}
\altaffiliation{Current address: Department of Physics and Astronomy, University of Tennessee, Knoxville, TN 37996, USA and Physics Division, Oak Ridge National Laboratory, Oak Ridge, TN 37831, USA.}
\affiliation{Centre de Saclay, IRFU/Service de Physique Nucl\'eaire, F-91191 Gif-sur-Yvette, France}

\author{Thomas Duguet}
\affiliation{Centre de Saclay, IRFU/Service de Physique Nucl\'eaire, F-91191 Gif-sur-Yvette, France}
\affiliation{National Superconducting Cyclotron Laboratory and Department of Physics and Astronomy, Michigan State University, East Lansing, MI 48824, USA}
\date{Updated \today}

\begin{abstract}
We present an extension of the pair coupled cluster doubles (p-CCD) method to quasiparticles and apply it to the attractive pairing Hamiltonian. Near the transition point where number symmetry gets spontaneously broken, the proposed BCS-based p-CCD method yields significantly better energies than existing methods when compared to exact results obtained via solution of the Richardson equations. The quasiparticle p-CCD method has a low computational cost of $\mathcal{O}(N^3)$ as a function of system size. This together with the high quality of results here demonstrated, points to considerable promise for the accurate description of strongly correlated systems with more realistic pairing interactions.
\end{abstract}
\maketitle

\section{Introduction}
The accurate description of weakly correlated fermionic systems of up to $\sim 10^2 - 10^3$ particles is by now fairly routine.  One can simply use the coupled cluster method\cite{CoesterKummel,BartlettShavitt,HagenCC}, where the correlated wave function is written as the exponential of an excitation operator acting on a mean-field reference determinant.  Truncating the excitation operator even at double excitations for systems with no more than two-body interactions recovers the vast majority of the correlation effects, and the perturbative inclusion of higher excitations recovers most of the rest.

The same cannot be said for systems of strongly correlated particles.  Exact diagonalization can of course be used when the number of particles is fairly small, but is too expensive for most systems of practical interest since the computational cost grows exponentially with system size.  Symmetry projected mean-field techniques\cite{BlaizotRipka,RingSchuck} and sophisticated multireference methods\cite{Roos} can be applied for somewhat larger systems, but eventually these too break down.

One way to extend the reach of diagonalization techniques is to restrict the wave function to be of low seniority, \textit{i.e.} to break only a small number of pairs.  With a suitably chosen pairing scheme, one can obtain a reasonable description of the strong correlations at a cost much reduced from that of the exact diagonalization.  For example, the cost of the zero-seniority diagonalization \cite{Volya}, which we will here call doubly occupied configuration interaction (DOCI) is roughly the square root of that of the exact diagonalization over the entire Hilbert space.  However, the cost still then scales exponentially with the size of the system, and even low-seniority diagonalizations are too expensive for most applications.

Recently, Refs. \onlinecite{Ayers1} and \onlinecite{Ayers2} introduced a method referred to as AP1roG and showed that it can efficiently account for strong correlations in molecular systems given the right single-particle levels and pairing scheme.  The AP1roG  is a zero-seniority wave function approach, and is formally equivalent to what we term pair coupled cluster doubles (p-CCD), which is a form of the coupled cluster method where the excitation operator is restricted to double excitations which do not break pairs.  They have shown that with an optimal mean-field reference and pairing scheme, p-CCD is remarkably close to the DOCI, despite having a computational cost which scales as $\mathcal{O}(N^3)$ where $N$ is some measure of the system size.  Thus, p-CCD seems to be a promising route toward the description of strongly correlated systems, provided that the correct reference determinant and pairing scheme can be easily found, and presuming that the correlation effects can be accurately captured by DOCI.

\begin{figure}[t]
\includegraphics[width=\columnwidth]{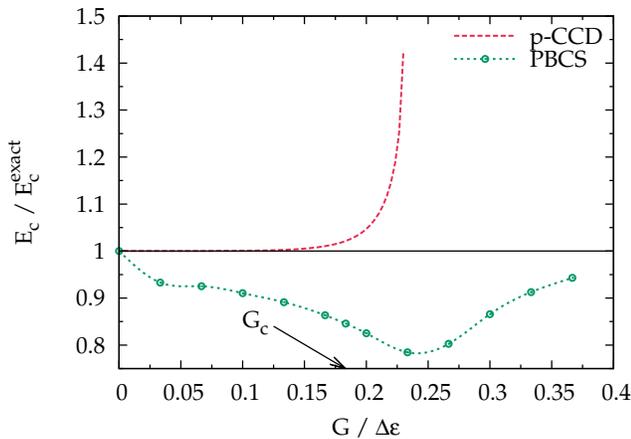}
\caption{Fraction of correlation energy recovered in the half-filled pairing Hamiltonian with 100 levels.  We show both the pair coupled cluster doubles and the number projected BCS.  A BCS solution emerges at the point we have labeled as $G_c$.  Past the point at which the coupled cluster curve is cut off, the method predicts a complex energy.
\label{Fig:AttractiveBreakDown}}
\end{figure}

The similarity between the p-CCD and DOCI is not, however, universal.  It has been shown \cite{JorgeBreakdown} that p-CCD and self-consistent RPA methods break down for the pairing Hamiltonian (see below) near the emergence of a number-broken BCS mean-field solution, as shown in Fig. \ref{Fig:AttractiveBreakDown}.  The pairing Hamiltonian studied in this work conserves seniority and therefore is solved exactly by DOCI; furthermore, p-CCD is equivalent to the full coupled cluster doubles (CCD) approach.  In spite of the strong reduction to the zero-seniority space, DOCI is limited to $\sim$36 levels at half filling.  For weakly attractive interactions (small $G$) CCD recovers the correlations very well.  However, CCD breaks down in a narrow region near $G_c$, the point at which there is a Hartree-Fock to BCS transition, and as the strength of the attractive interaction continues to increase, we find a complex correlation energy.  The number-projected BCS (PBCS), in contrast, captures the essential large-$G$ physics, and is very accurate for sizes where DOCI is applicable \cite{Sandulescu}.  In fact, the BCS itself picks up much of the correlation effects.  This suggests that we try a BCS-based quasiparticle CCD; letting the mean-field describe the most important energetic features of the strong correlations frees the coupled cluster method to focus on the description of the remaining weaker correlations, for which it excels.

The remainder of this paper is organized as follows.  Section \ref{Sec:Hamiltonian} discusses the pairing Hamiltonian which we will be interested in solving with the quasiparticle coupled cluster theory presented in Sec. \ref{Sec:CCTheory}.  We provide several results in Sec. \ref{Sec:Results} to illustrate the predictive ability of our methodology, and conclude with a brief discussion in Sec. \ref{Sec:Discussion}.

\section{The Pairing Hamiltonian
\label{Sec:Hamiltonian}}
The pairing or reduced BCS Hamiltonian can be written as
\begin{equation}
H = \sum_i \left(\epsilon_i - \lambda\right) N_i - G \sum_{ij} P_i^\dagger \, P_j.
\label{Eqn:PairHam}
\end{equation}
Here, $\lambda$ is the chemical potential, the $\epsilon_i$ are single-particle energy levels, and $G$ is the interaction strength.  We have defined pair operators
\begin{subequations}
\begin{align}
N_i &= a_{i_\uparrow}^\dagger \, a_{i_\uparrow}^{} + a_{i_\downarrow}^\dagger \, a_{i_\downarrow}^{},
\\
P_i^\dagger &= a_{i_\uparrow}^\dagger \, a_{i_\downarrow}^\dagger.
\end{align}
\label{Eqn:DefNPhys}
\end{subequations}
These operators satisfy an $SU(2)$ algebra, with
\begin{subequations}
\begin{align}
[P_i,P_j^\dagger] &= \delta_{ij} \, \left(1 - N_i\right),
\\
[N_i,P_j] &= -2 \, \delta_{ij} \, P_j,
\\
[N_i,P_j^\dagger] &= 2 \, \delta_{ij} \, P_J^\dagger.
\end{align}
\label{SU2}
\end{subequations}
For simplicity, we will take the single-particle levels to be equally spaced, so that $\epsilon_p = p \, \Delta \epsilon$ where $\Delta \epsilon$ is the level spacing.

Originally developed to phenomenologically describe superconductivity in solids\cite{Bardeen}, BCS was soon realized to explain the large gaps observed in even-even nuclei\cite{Pines} as well.  However, nuclei are finite systems and the superconducting correlations should be strongly influenced by the finite effects.  Since then, and up to the present, number projection \cite{Pradal}  and in general symmetry restoration in the BCS and Hartree-Fock-Bogoliubov approximations have been important issues in nuclear structure and more recently in quantum chemistry.

At the beginning of the sixties Richardson provided an exact solution for the reduced BCS Hamiltonian of Eqn. \ref{Eqn:PairHam} \cite{Richardson1,Richardson4}.  In spite of the importance of his exact solution, this work did not have much impact in nuclear physics, with just a few exceptions. Later on, his exact solution was rediscovered in the framework of ultrasmall superconducting grains \cite{Sierra} where BCS and number-projected BCS \cite{Braun} were unable to appropriately describe the crossover from superconductivity to a normal metal as a function of the grain size \cite{Dukelsky}.  The ability to access the exact solutions for systems far beyond the reach of diagonalization techniques makes the model particularly appealing for our purposes.  Presumably, if we can develop computationally efficient techniques which can capture the basic physics of the pairing Hamiltonian, we can extend those techniques to systems with more realistic pairing interactions for which the exact solutions are unavailable.  It is with this purpose in mind that we explore the quasiparticle coupled cluster method.  Accordingly, while we will apply the methodology presented below to the pairing Hamiltonian of Eqn. \ref{Eqn:PairHam}, we will keep the formulation general enough that it can be applied immediately to \textit{any} two-body Hamiltonian expressible in terms of the operators $N$, $P^\dagger$, and $P$.  

We should point out that whenever the Hamiltonian has a natural pairing scheme so that seniority is a good quantum number, the coupled-cluster method reduces to the pair coupled-cluster approach.  For Hamiltonians in which seniority is not a symmetry, it may be desirable to use a more general Hartree--Fock--Bogoliubov-based coupled cluster technique, the equations for which will be presented in due time\cite{Duguet}.  Our BCS-based coupled cluster is a special case of this more general technique.  The quasiparticle perturbation theory of Lacroix and Gambacurta \cite{LaCroixGambacurta} is related, though we do not consider augmenting the coupled cluster approach with a subsequent number pojection.

\section{Quasiparticle Coupled Cluster Theory
\label{Sec:CCTheory}}
For sufficiently strong $G$, the pairing Hamiltonian develops a BCS solution with quasiparticle operators defined by
\begin{subequations}
\begin{align}
a_{i_\uparrow}^\dagger &= u_i \, \alpha_{i_\uparrow}^\dagger + v_i \, \alpha_{i_\downarrow}^{},
\\
a_{i_\downarrow}^\dagger &= u_i \, \alpha_{i_\downarrow}^\dagger - v_i \, \alpha_{i_\uparrow}^{}.
\end{align}
\label{Eqn:QPTransformation}
\end{subequations}
The Hamiltonian can be expressed in this quasiparticle basis as
\begin{align}
H &= E_0
   + \sum_i H_i^{1,1} \N_i
   + \sum_i \left(H_i^{0,2} \, \P_i^\dagger + H_i^{2,0} \, \P_i\right)
\\
  &+ \sum_{ij} H_{ij}^{2,2} \N_i \, \N_j
   + \sum_{ij} \tilde{H}_{ij}^{2,2} \, \P_i^\dagger \, \P_j
\nonumber
\\
  &+ \sum_{ij} \left(H_{ij}^{0,4} \, \P_i^\dagger \, \P_j^\dagger + H_{ij}^{4,0} \, \P_i \, \P_j\right)
\nonumber
\\
  &+ \sum_{ij} \left(H_{ij}^{1,3} \, \P_i^\dagger \, \N_j + H_{ij}^{3,1} \, \N_j \, \P_i \right)
\nonumber
\end{align}
where the number and pair operators are
\begin{subequations}
\begin{align}
\N_i &= \alpha_{i_\uparrow}^\dagger \, \alpha_{i_\uparrow}^{} + \alpha_{i_\downarrow}^\dagger \, \alpha_{i_\downarrow}^{},
\\
\P_i^\dagger &= \alpha_{i_\uparrow}^\dagger \, \alpha_{i_\downarrow}^\dagger
\end{align}
\end{subequations}
and obey commutation relations analagous to those in Eqn. \ref{SU2}.  Most of the matrix elements of the Hamiltonian are symmetric, but note the order of indices on the non-symmetric $H^{1,3}$ and $H^{3,1}$.  The diagonal entries of $H^{4,0}$ and $H^{0,4}$ are undefined since $\P_p^\dagger \, \P_p^\dagger = 0$, and for convenience we take them to vanish.  Expressions for the individual matrix elements are provided in the appendix.

Having defined a mean-field vacuum, explicit correlations are added via the coupled cluster (CC) method.  Briefly, in CC theory, the correlated wave function $|\Psi\rangle$ is written as
\begin{equation}
|\Psi\rangle = \mathrm{e}^T |0\rangle
\end{equation}
where $|0\rangle$ is the mean-field reference and $T$ is an excitation operator which we may generically write as
\begin{equation}
T = \sum_Q T_Q \mathcal{A}_Q^\dagger.
\end{equation}
Here, $\mathcal{A}_Q^\dagger$ creates excitations out of a mean-field reference, and $T_Q$ is the amplitude of the excitation.  The CC wave function ansatz is inserted into the Schr\"odinger equation and the energy $E_{CC}$ and amplitudes $T_Q$ are obtained projectively from
\begin{subequations}
\begin{align}
E_{CC} &= \langle 0 | \mathrm{e}^{-T} \, H \, \mathrm{e}^T | 0\rangle,
\label{CCEnergy}
\\
0 &= \langle 0 | \mathcal{A}_Q \, \mathrm{e}^{-T} \, H \, \mathrm{e}^T | 0\rangle.
\end{align}
\label{BasicCCEquations}
\end{subequations}

Conventionally, CC theory uses a number-conserving mean-field and the excitation operators $\mathcal{A}_Q^\dagger$ are just particle-hole excitations $a_a^\dagger \, a_i$, $a_a^\dagger \, a_i \, a_b^\dagger \, a_j$, \textit{etc.}, where indices $i$ and $j$ ($a$ and $b$) correspond to levels occupied (empty) in the mean-field reference.  In our case, we wish to use BCS as a reference instead, and we write the excitation operator $T$ as
\begin{equation}
T = \frac{1}{2} \sum_{pq} T_{pq} \, \P_p^\dagger \, \P_q^\dagger.
\label{CCAnsatz}
\end{equation}
The matrix of coefficients $T_{pq}$ is symmetric, and as with $H^{4,0}$ and $H^{0,4}$, we choose the diagonal entries to vanish.  The cluster operator we have introduced is the simplest useful form for $T$ and is a generalization to the BCS case of p-CCD, though we remind the reader that for the pairing Hamiltonian, p-CCD and the full CCD model are identical.  We note that coupled cluster theory has been formulated in a quasiparticle basis before \cite{StolarczykMonkhorst}, but with a restriction that the wave function does not break number symmetry.

Having defined the cluster operator $T$, the CC energy of Eqn. \ref{CCEnergy} is given by
\begin{equation}
E_{CC} = E_0 + \sum_{pq} H^{4,0}_{pq} \, T_{pq}
\end{equation}
and the amplitude equations are
\begin{widetext}
\begin{subequations}
\begin{align}
0
 &= \langle 0| \P_p \, \P_q \, \mathrm{e}^{-T} \, H \, \mathrm{e}^T | 0 \rangle
\\
 &= 2 \, H_{pq}^{0,4}
  + 2 \, T_{pq} \, \left(H_p^{1,1} + H_q^{1,1}\right)
  + 4 \, T_{pq} \, \left(H_{pp}^{2,2} + 2 \, H_{pq}^{2,2} + H_{qq}^{2,2} \right)
  + \sum_r \tilde{H}^{2,2}_{pr} \, T_{qr}
  + \sum_r \tilde{H}^{2,2}_{qr} \, T_{pr}
\label{T2Eqns}
\\
 &+ 2 \, \sum_{rs} H_{rs}^{4,0} \, T_{pr} \, T_{qs}
  - 4 \, T_{pq} \, \left(\sum_r H_{pr}^{4,0} \, T_{pr} + \sum_r H_{qr}^{4,0} \, T_{qr}\right)
  + 4 \, H_{pq}^{4,0} \, T_{pq}^2.
\nonumber
\end{align}
\end{subequations}
\end{widetext}
We will refer to this BCS-based p-CCD as BCS p-CCD, to distinguish it from the Hartree--Fock-based p-CCD which we will henceforth refer to as HF p-CCD for clarity.

It is well known that the CCD energy and wave function can be rather sensitive to the choice of mean-field reference.  This can be mitigated by explicitly including single excitations which act as a Thouless transformation to relax the reference.  Alternatively, one could adjust the reference such that single excitations self-consistently vanish using what we refer to as the Brueckner determinant\cite{Brueckner}.  We prefer the latter approach, and supplement the amplitude equations of Eqn. \ref{T2Eqns} by
\begin{subequations}
\begin{align}
0 &= \langle 0| \P_p \, \mathrm{e}^{-T} \, H \, \mathrm{e}^T |0 \rangle
\\
  &= H_p^{0,2} + \sum_q \left(H_q^{2,0} + 2 \, H_{qp}^{3,1}\right) \, T_{pq}
\label{T1Eqn}
\end{align}
\end{subequations}
which we solve by modifying the $u_p$ and $v_p$ parameters defining the quasiparticle transformation.  Explicitly, we have a modified BCS equation:
\begin{equation}
2 \, \xi_p \, u_p \, v_p - \Delta \, \left(u_p^2 - v_p^2\right) + c_p = 0
\end{equation}
with
\begin{subequations}
\begin{align}
\xi_p &= \epsilon_p - \lambda - G \, v_p^2 + G \, \sum_q \left(u_q^2 - v_q^2\right) \, T_{pq},
\\
\Delta &= G \, \sum_q u_q \, v_q
\\
c_p &= \sum_q H^{2,0}_q \, T_{pq}.
\end{align}
\end{subequations}
We solve this modified BCS equation with
\begin{equation}
v_p^2 = \frac{\xi_p^2 + \Delta^2 - c_p \, \Delta - \xi_p \, \sqrt{\xi_p^2 + \Delta^2 - c_p^2}}{2 \left(\xi_p^2 + \Delta^2\right)}
\end{equation}
such that we obtain the BCS amplitudes for $T \to 0$.  When we supplement our BCS p-CCD with the Brueckner condition of Eqn. \ref{T1Eqn} we will refer to the method as BCS p-BCCD (for BCS pair Brueckner coupled cluster doubles).  The quasiparticle determinant defined by the Brueckner condition will be called the BCS-Brueckner determinant.

We should note that when the BCS reduces to Hartree-Fock, the CC ansatz of Eqn. \ref{CCAnsatz} becomes
\begin{equation}
T \to \sum_{ai} T_{ai} \, P_a^\dagger \, P_i + \frac{1}{2} \, \sum_{ij} T_{ij} \, P_i \, P_j + \frac{1}{2} \sum_{ab} T_{ab} \, P_a^\dagger \, P_b^\dagger
\end{equation}
where we recall that indices $i$ and $j$ ($a$ and $b$) refer to orbitals occupied (unoccupied) in the Hartree-Fock determinant.  While the first term is HF p-CCD, the other two terms would appear to break number symmetry.  This is not the case, however, because the amplitudes $T_{ij}$ and $T_{ab}$ vanish.  More precisely, if a set of amplitude $T_{ai}$ solve the HF p-CCD equations, then they, together with $T_{ij} = T_{ab} = 0$, also solve the BCS p-CCD equations.  Thus, the Hartree-Fock limit of our BCS-based p-CCD is just Hartree--Fock-based p-CCD.  The BCS-Brueckner determinant in this limit is the same as the Hartree-Fock determinant.

Note finally that the chemical potential which yields the correct average particle number for the BCS wave function does not, in general, yield the right average particle number from the coupled cluster wave function.  We circumvent this problem here by working primarily at half filling, for which our reduced BCS Hamiltonian has a particle-hole symmetry\cite{ParticleHoleSymmetry}; this symmetry means that the BCS and BCS p-CCD wave functions both contain the correct number of particles on average with the BCS chemical potential.

\begin{figure}[t]
\includegraphics[width=\columnwidth]{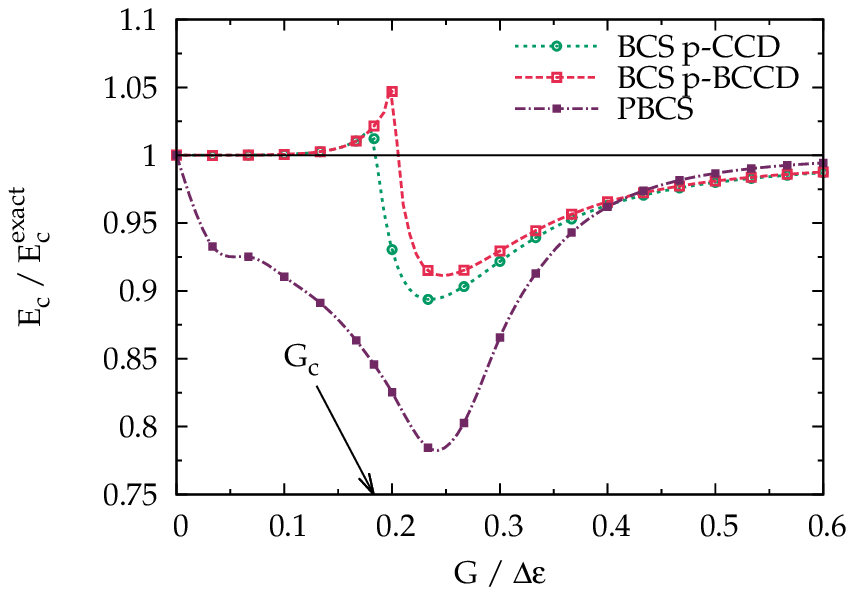}
\caption{Fraction of correlation energy recovered in the half-filled pairing Hamiltonian with 100 levels.  BCS p-CCD and BCS p-BCCD refer to p-CCD based on a BCS or on a BCS-Brueckner reference, respectively.
\label{Fig:Attractive}}
\end{figure}

\section{Results
\label{Sec:Results}}
We have already seen that HF p-CCD breaks down in a narrow region around the Hartree-Fock to BCS transition.  As Fig. \ref{Fig:Attractive} shows, the same is not true of BCS p-CCD.  While the results in the immediate vicinity of $G_c$ degrade somewhat, the BCS p-CCD results improve again as $G$ becomes large and the BCS is able to recover the energetic effects of the strong correlations in the system.  We note that for very small systems, the projected BCS is more accurate than is the BCS p-CCD (as we shall see later).  Recall, however, that for sufficiently large systems the projected BCS returns the same correlation energy per particle as does BCS itself\cite{staroverov2002}.  In contrast, the BCS p-CCD will offer a non-negligible improvement upon BCS for any system size, because it is size-extensive (\textit{i.e.} the correlation energy per particle approaches a constant for large particle number).  For the 100-level pairing Hamiltonian, we see that the BCS p-CCD is generally superior to PBCS except for very large $G$.

\begin{figure}[t]
\includegraphics[width=\columnwidth]{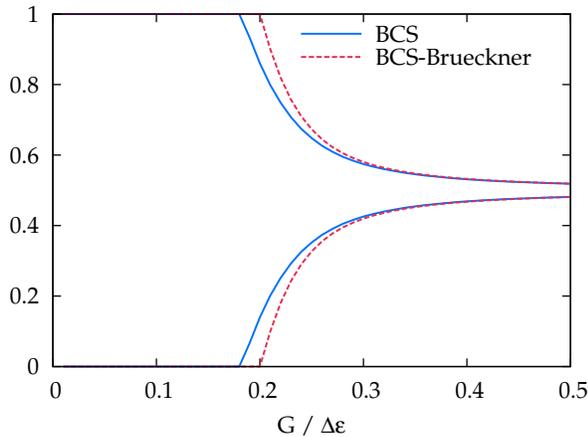}
\caption{Occupation numbers from the BCS and BCS-Brueckner determinants as a function of $G$ for the half-filled pairing Hamiltonian with 100 levels.  We show the two occupation numbers which bracket the Fermi level at $G=0$.
\label{Fig:OccNosMeanField}}
\end{figure}

A curious feature of Fig. \ref{Fig:Attractive} is that the BCS--Bruckner-based p-CCD splits off from the BCS p-CCD near the Hartree-Fock to BCS transition.  This is simply because the Brueckner procedure defers the onset of number symmetry breaking, much like it delays the onset of spin symmetry breaking as one stretches a chemical bond\cite{BruecknerSymm1,BruecknerSymm2,BruecknerSymm3,BruecknerSymm4}.  One can see that by examining the occupation numbers $\tfrac{1}{2} \langle 0| N_p | 0\rangle$ of the BCS and BCS-Brueckner reference determinants.  As we see in Fig. \ref{Fig:OccNosMeanField}, the BCS-Brueckner determinant breaks number symmetry at larger $G$ than does the BCS itself.

A second feature of interest may be the behavior of the BCS p-CCD at the Hartree-Fock to BCS transition.  While the mean-field transition is second-order in nature, the p-CCD appears to have a first-order transition (\textit{i.e.} the energy is continuous as a function of $G$ but its derivative is not).  One sees analogous behavior in molecule dissociations, where CCD has a derivative discontinuity at the point of spin symmetry breaking.  The inclusion of explicit single excitations remedies this defect, and we expect it would do likewise here.

Thus far, we have shown that the BCS p-CCD can capture the energetic effects of the correlation beyond that provided by BCS.  Perhaps more interesting is to use the BCS p-CCD to evaluate properties.  The evaluation of expectation values within the coupled cluster framework is complicated somewhat by the fact that the coupled cluster method is non-variational.  One can eliminate the need to evaluate the dependence of the cluster amplitudes $T_{pq}$ on Hamiltonian parameters through the use of linear response techniques\cite{BartlettShavitt}.  Simpler, however, is to differentiate the coupled cluster energy with respect to those parameters.  Thus, for example, the occupation probability $\langle N_p \rangle$ can be evaluated as $\tfrac{\mathrm{d}E}{\mathrm{d}\epsilon_p}$.  We take the latter approach in this work, though in the immediate vicinity of the Hartree-Fock to BCS transition this numerical derivative cannot be evaluated reliably because the coupled cluster wave function changes character abruptly at $G_c$.

\begin{figure}[t]
\includegraphics[width=\columnwidth]{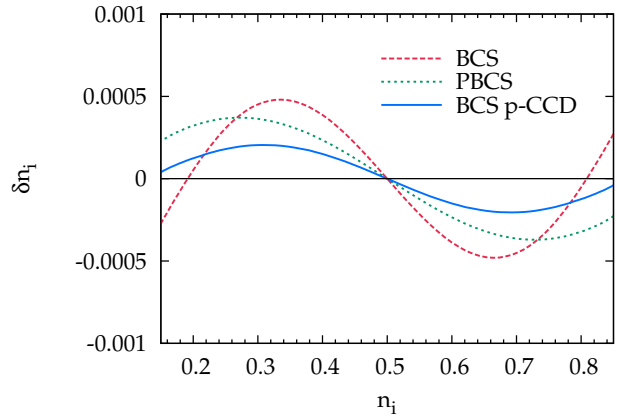}
\caption{Deviations from the exact occupation number for the half-filled pairing Hamiltonian with 100 levels and $G / \Delta \epsilon$ = 1.
\label{Fig:OccNosError}}
\end{figure}

Figure \ref{Fig:OccNosError} shows the error in the single-particle occupation probabilities from BCS and from BCS p-CCD.  We can compute these by simply taking the numerical derivative of the BCS p-CCD energy with respect to the single-particle energies $\epsilon_p$.  We see that the BCS p-CCD reduces the errors in the occupation probabilities by roughly a factor of two compared to BCS.  A consequence of halving the errors in the occupation probabilities is that we would expect to roughly halve the error in \textit{any} one-particle property. 

\begin{figure}[t]
\includegraphics[width=\columnwidth]{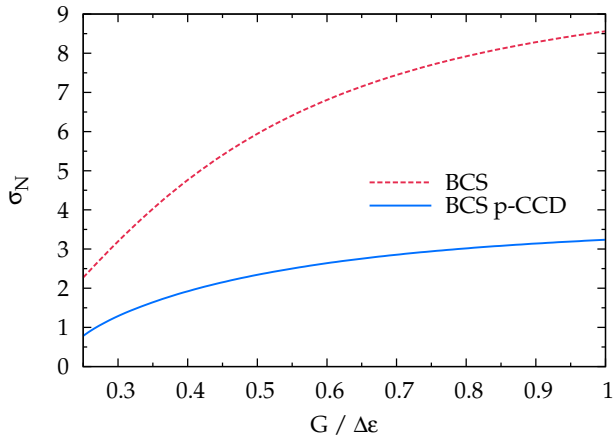}
\caption{Fluctuations in particle number in the BCS and BCS p-CCD wave functions for the half-filled pairing Hamiltonian with 100 levels.
\label{Fig:TotalNumber}}
\end{figure}

In Fig. \ref{Fig:TotalNumber}, we show the deviation in particle number, given schematically by
\begin{equation}
\sigma_N^2 = \langle N^2\rangle - \langle N \rangle^2.
\end{equation}
For the PBCS and exact wave functions, this quantity is of course exactly zero; again, the BCS p-CCD cuts the error approximately in half.

\begin{figure}[t]
\includegraphics[width=\columnwidth]{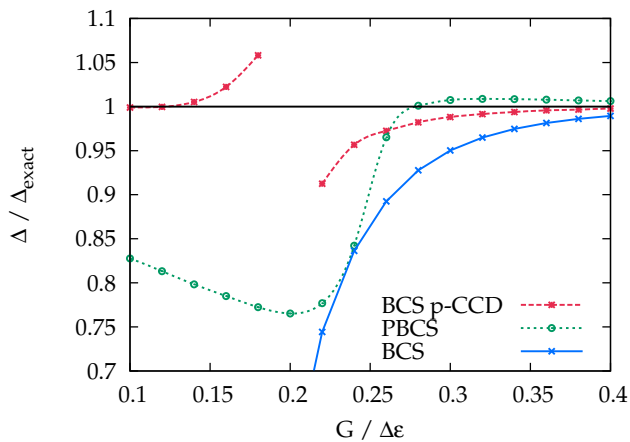}
\caption{Pairing parameter $\Delta_c$ defined in Eqn. \ref{Eqn:DeltaC} for the half-filled pairing Hamiltonian with 100 levels.
\label{Fig:Gaps}}
\end{figure}

As a final example of the evaluation of properties, we consider a generalization of the BCS superconducting gap $\Delta$ to the case of correlated wave functions.  We use the pairing parameter proposed in Refs. \onlinecite{Tichy} and \onlinecite{Braun}:
\begin{equation}
\Delta_c = G \sum_p C_p
\label{Eqn:DeltaC}
\end{equation}
where
\begin{equation}
C_p^2
 = \langle P_p^\dagger \, P_p \rangle - \frac{1}{4} \langle N_p \rangle^2
 = \langle n_{p_\uparrow} n_{p_\downarrow} \rangle - \frac{1}{4} \langle N_p \rangle^2.
\end{equation}
Here, $n_{p_\sigma} = a_{p_\sigma}^\dagger a_{p_\sigma}$.  In the BCS case, this pairing parameter $\Delta_c$ reduces to the usual superconducting gap $\Delta = G \sum u_p v_p$.  While the BCS gap vanishes when number symmetry is not broken, the parameter $\Delta_c$ will be in general non-zero for correlated wave functions.  Figure \ref{Fig:Gaps} shows results for this pairing parameter.  For both the weakly attractive and strongly attractive cases, the coupled cluster appears to better predict this order parameter than does projected BCS, though for smaller systems projected BCS is somewhat more accurate in the intermediate coupling regime (data not shown, but see also Ref. \onlinecite{Sandulescu}).

\begin{figure}[t]
\includegraphics[width=\columnwidth]{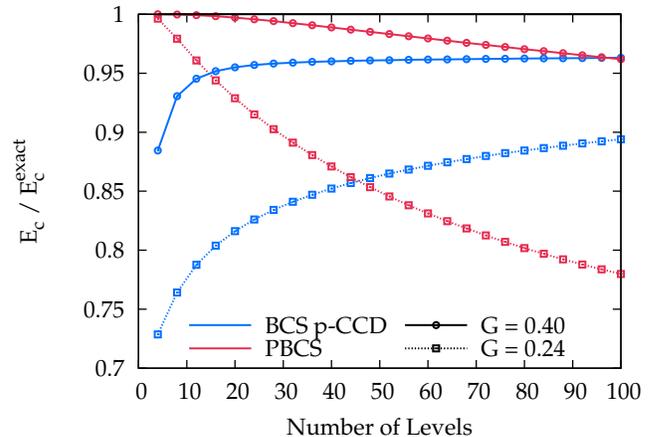}
\caption{Fraction of the correlation energy recovered in half-filled pairing Hamiltonians as a function of the number of levels.  We have scaled the interaction strength $G$ to keep a roughly constant BCS gap and have used $G = 0.24$ and $G = 0.40$ for 100 levels.
\label{Fig:EvsL}}
\end{figure}

\begin{figure}[t]
\includegraphics[width=\columnwidth]{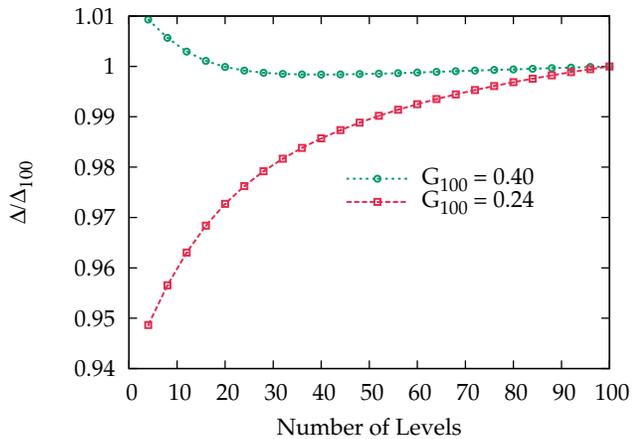}
\caption{BCS gap $\Delta$ as a function of the number of levels with pairing strength $G$ scaled according to Eqn. \ref{ScaleDelta}.
\label{Fig:DeltaVsL}}
\end{figure}

All our results thus far have been generated for pairing Hamiltonians with 100 levels.  Figure \ref{Fig:EvsL} shows how the coupled cluster method performs as we change the number of levels while keeping the filling fraction constant.  Because for a given value of $G$ and a given filling fraction, increasing the number of levels increases the degree of correlation, we have chosen to scale $G$ in order to keep the BCS gap $\Delta$ roughly constant (see Fig. \ref{Fig:DeltaVsL}), using\cite{ConstantBCSGap}
\begin{equation}
\frac{1}{L} \, \sinh\left(\frac{1}{G}\right) = \mathrm{constant}
\label{ScaleDelta}
\end{equation}
where $L$ is the number of levels; we have chosen the constant such that $G = 0.24$ and $G = 0.40$ for the 100-level problem, as these correspond to values of $G$ for which PBCS and BCS p-CCD are least accurate ($G = 0.24$) and for which PBCS and BCS p-CCD are comparable in accuracy ($G =0.40)$.  We see clearly that PBCS degrades in accuracy as the system size increases, while the relative error from the CC method actually decreases as we increase the number of levels.  

\begin{figure}[t]
\includegraphics[width=\columnwidth]{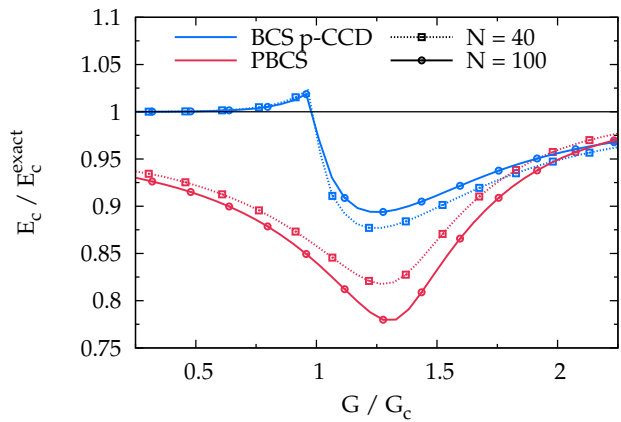}
\caption{Fraction of correlation recovered in the pairing Hamiltonian with 100 levels.  Solid lines denote half filling while dotted lines indicate 20\% filling.  The interaction strength $G$ is expressed in units of $G_c$, the value of $G$ for which the Hartree-Fock to BCS transition occurs.
\label{Fig:Filling20}}
\end{figure}

While we have focused on the half-filling case for simplicity, the coupled cluster method is general.  We may use
\begin{equation}
\langle N \rangle = \frac{\mathrm{d}}{\mathrm{d}\lambda} \langle 0| \mathrm{e}^{-T}\, \left(H + \lambda \, N\right) \, \mathrm{e}^T |0\rangle
\end{equation}
to obtain the average number of particles in the coupled-cluster wave function and then adjust the chemical potential to force this average to be correct; note that the BCS wave function therefore has the wrong average particle number, though the coupled cluster and BCS chemical potentials are generally quite similar.  Figure \ref{Fig:Filling20} shows the results at half-filling ($N=100$) and at 20\% filling ($N=40$) for the 100-level pairing Hamiltonian; we have written the interaction strength $G$ as a multiple of $G_c$, where the Hartree-Fock to BCS transition occurs.  Clearly, results for the 20\% filling case are qualitatively very similar to what we see at half-filling, though we note that here the coupled cluster results are somewhat better at half filling while the PBCS results are somewhat better away from half filling.

\begin{figure}[t]
\includegraphics[width=\columnwidth]{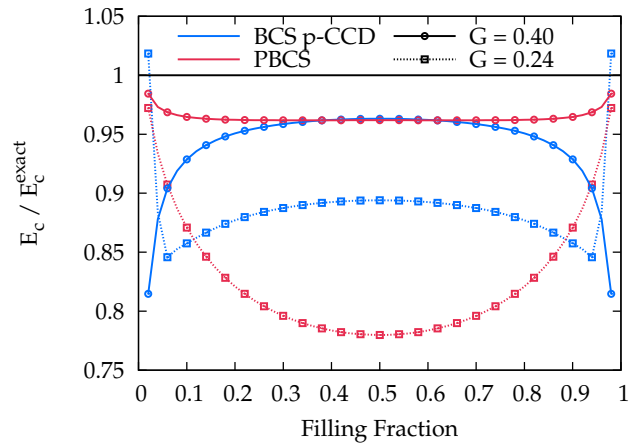}
\caption{Fraction of correlation energy recovered in the pairing Hamiltonian with 100 levels plotted against the filling fraction.
\label{Fig:EvsN}}
\end{figure}

To further illustrate the dependence of the coupled cluster method on particle number, Fig. \ref{Fig:EvsN} shows the fraction of correlation energy recovered in the 100-level pairing Hamiltonian as a function of the filling fraction for two different values of $G$ ($G = 0.24$ and $G = 0.40$, for reasons discussed above).  For very small numbers of particles or holes, preserving number symmetry appears to be essential -- and note that at $G = 0.24$, the BCS does not break number symmetry for small filling and consequently neither does BCS p-CCD.  Away from these two extremes, however, the BCS p-CCD performs very consistently.  We should note that while the BCS p-CCD appears to break down rather badly for $G = 0.40$ and small filling fractions, this is to some extent illusory in that the error in the total correlation energy is very small for small or large filling.

\section{Discussion
\label{Sec:Discussion}}
Because the pairing Hamiltonian has a simple exact solution available, it provides a very useful model for the testing of approximate solutions of the Sch\"odinger equation.  While coupled cluster theory has generally provided very accurate wave functions even in its simplest form, it breaks down badly near the Hartree-Fock to BCS transition in the attractive pairing Hamiltonian.  The success of BCS and of number-projected BCS for this problem suggest that a quasiparticle coupled cluster ansatz based on the BCS vacuum should succeed where Hartree--Fock-based coupled cluster fails.

Indeed, while our BCS p-CCD reduces to the HF p-CCD for repulsive interactions, it also accurately describes the attractive interactions in the pairing Hamiltonian.  Unlike exact diagonalization, the computational scaling with system size is very mild.  Unlike symmetry projected mean-field methods, the correlation energy per particle approaches a non-zero constant for large system sizes.  The BCS p-CCD not only describes the energetic effects of strong correlations but also accounts for the effects of these correlations on other properties.  Adjusting the BCS reference to be self-consistent in the presence of correlations yields BCS p-BCCD, which defers the onset of symmetry breaking and offers an even better description of the attractive pairing Hamiltonian than does the BCS p-CCD itself.  The inclusion of explicit higher-order correlation effects is also possible and fairly straightforward; doing so would presumably further increase the accuracy of the approach.  We would thus argue that BCS-based coupled cluster methods show considerable promise for the description of strongly correlated systems with more realistic pairing interactions.

\section{Acknowledgments}
This work was supported by the Department of Energy, Office of Basic Energy Sciences, Grant No. DE-FG02-09ER16053, and by the Spanish MINECO under grand FIS2012-34479.  G.E.S. is a Welch Foundation Chair (C-0036).  A. S. acknowledges support from Espace de Structure et R\'eaction Nucl\'eaire Th\'eorique (ESNT).  We thank Peter Schuck for helpful comments.

\appendix
\section{Quasiparticle Matrix Elements}
Recall that the pairing Hamiltonian in the bare Fermion basis is
\begin{equation}
H = \sum_i \left(\epsilon_i - \lambda\right) N_i - G \sum_{ij} P_i^\dagger \, P_j.
\end{equation}
After the quasiparticle transformation
\begin{subequations}
\begin{align}
a_{i_\uparrow}^\dagger &= u_i \, \alpha_{i_\uparrow}^\dagger + v_i \, \alpha_{i_\downarrow}^{},
\\
a_{i_\downarrow}^\dagger &= u_i \, \alpha_{i_\downarrow}^\dagger - v_i \, \alpha_{i_\uparrow}^{},
\end{align}
\end{subequations}
it is equivalently given by
\begin{align}
H &= E_0
   + \sum_i H_i^{1,1} \N_i
   + \sum_i \left(H_i^{0,2} \, \P_i^\dagger + H_i^{2,0} \, \P_i\right)
\\
  &+ \sum_{ij} H_{ij}^{2,2} \N_i \, \N_j
   + \sum_{ij} \tilde{H}_{ij}^{2,2} \, \P_i^\dagger \, \P_j
\nonumber
\\
  &+ \sum_{ij} \left(H_{ij}^{0,4} \, \P_i^\dagger \, \P_j^\dagger + H_{ij}^{4,0} \, \P_i \, \P_j\right)
\nonumber
\\
  &+ \sum_{ij} \left(H_{ij}^{1,3} \, \P_i^\dagger \, \N_j + H_{ij}^{3,1} \, \N_j \, \P_i \right).
\nonumber
\end{align}
We again emphasize the order of indices on $H^{1,3}$ and $H^{3,1}$.

In terms of the Fock matrix element
\begin{equation}
\mathcal{F}_p = \epsilon_p - \lambda - G \, v_p^2
\end{equation}
and the pairing field
\begin{equation}
\Delta = G \, \sum_p u_p \, v_p,
\end{equation}
the quasiparticle Hamiltonian matrix elements are
\begin{subequations}
\begin{align}
E_0 &= \sum_p v_p^2 \, \left(2 \, \epsilon_p - G \, v_p^2\right) - \frac{\Delta^2}{G},
\\
H_i^{1,1}
 &= \left(\epsilon_i - \lambda\right) \, \left(u_i^2 - v_i^2\right)
\\
 &+ 2 \, \Delta \, u_i \, v_i + G \, v_i^4,
\nonumber
\\
H_i^{2,0} &= H_i^{0,2} =2 \, \mathcal{F}_i \, u_i \, v_i - \Delta \, \left(u_i^2 - v_i^2\right),
\nonumber
\\
H_{ij}^{2,2} &= - G \, u_i \, v_i \, u_j \, v_j,
\\
\tilde{H}_{ij}^{2,2} &= - G \, \left(u_i^2 \, u_j^2 + v_i^2 \, v_j^2 \right),
\\
H_{ij}^{3,1} &= H_{ij}^{1,3} = G \, \left(u_i^2 - v_i^2 \right) \, u_j \, v_j,
\\
H_{ij}^{4,0} &= H_{ij}^{0,4} = G \, \frac{u_i^2 \, v_j^2 + v_i^2 \, u_j^2}{2}.
\end{align}
\end{subequations}
As noted earlier, we have taken the diagonal parts of $H^{4,0}$ and $H^{0,4}$ to vanish for convenience.

\bibliography{HFBCC}

\begin{thebibliography}{31}%
\makeatletter
\providecommand \@ifxundefined [1]{%
 \@ifx{#1\undefined}
}%
\providecommand \@ifnum [1]{%
 \ifnum #1\expandafter \@firstoftwo
 \else \expandafter \@secondoftwo
 \fi
}%
\providecommand \@ifx [1]{%
 \ifx #1\expandafter \@firstoftwo
 \else \expandafter \@secondoftwo
 \fi
}%
\providecommand \natexlab [1]{#1}%
\providecommand \enquote  [1]{``#1''}%
\providecommand \bibnamefont  [1]{#1}%
\providecommand \bibfnamefont [1]{#1}%
\providecommand \citenamefont [1]{#1}%
\providecommand \href@noop [0]{\@secondoftwo}%
\providecommand \href [0]{\begingroup \@sanitize@url \@href}%
\providecommand \@href[1]{\@@startlink{#1}\@@href}%
\providecommand \@@href[1]{\endgroup#1\@@endlink}%
\providecommand \@sanitize@url [0]{\catcode `\\12\catcode `\$12\catcode
  `\&12\catcode `\#12\catcode `\^12\catcode `\_12\catcode `\%12\relax}%
\providecommand \@@startlink[1]{}%
\providecommand \@@endlink[0]{}%
\providecommand \url  [0]{\begingroup\@sanitize@url \@url }%
\providecommand \@url [1]{\endgroup\@href {#1}{\urlprefix }}%
\providecommand \urlprefix  [0]{URL }%
\providecommand \Eprint [0]{\href }%
\providecommand \doibase [0]{http://dx.doi.org/}%
\providecommand \selectlanguage [0]{\@gobble}%
\providecommand \bibinfo  [0]{\@secondoftwo}%
\providecommand \bibfield  [0]{\@secondoftwo}%
\providecommand \translation [1]{[#1]}%
\providecommand \BibitemOpen [0]{}%
\providecommand \bibitemStop [0]{}%
\providecommand \bibitemNoStop [0]{.\EOS\space}%
\providecommand \EOS [0]{\spacefactor3000\relax}%
\providecommand \BibitemShut  [1]{\csname bibitem#1\endcsname}%
\let\auto@bib@innerbib\@empty
\bibitem [{\citenamefont {Coester}\ and\ \citenamefont
  {K\"ummel}(1960)}]{CoesterKummel}%
  \BibitemOpen
  \bibfield  {author} {\bibinfo {author} {\bibfnamefont {F.}~\bibnamefont
  {Coester}}\ and\ \bibinfo {author} {\bibfnamefont {H.}~\bibnamefont
  {K\"ummel}},\ }\href@noop {} {\bibfield  {journal} {\bibinfo  {journal}
  {Nuclear Physics}\ }\textbf {\bibinfo {volume} {17}},\ \bibinfo {pages} {477}
  (\bibinfo {year} {1960})}\BibitemShut {NoStop}%
\bibitem [{\citenamefont {Shavitt}\ and\ \citenamefont
  {Bartlett}(2009)}]{BartlettShavitt}%
  \BibitemOpen
  \bibfield  {author} {\bibinfo {author} {\bibfnamefont {I.}~\bibnamefont
  {Shavitt}}\ and\ \bibinfo {author} {\bibfnamefont {R.~J.}\ \bibnamefont
  {Bartlett}},\ }\href@noop {} {\emph {\bibinfo {title} {Many-Body Methods in
  Chemistry and Physics}}}\ (\bibinfo  {publisher} {Cambridge University
  Press},\ \bibinfo {year} {2009})\BibitemShut {NoStop}%
\bibitem [{\citenamefont {Hagen}\ \emph {et~al.}()\citenamefont {Hagen},
  \citenamefont {Papenbrock}, \citenamefont {Hjorth-{J}ensen},\ and\
  \citenamefont {Dean}}]{HagenCC}%
  \BibitemOpen
  \bibfield  {author} {\bibinfo {author} {\bibfnamefont {G.}~\bibnamefont
  {Hagen}}, \bibinfo {author} {\bibfnamefont {T.}~\bibnamefont {Papenbrock}},
  \bibinfo {author} {\bibfnamefont {M.}~\bibnamefont {Hjorth-{J}ensen}}, \ and\
  \bibinfo {author} {\bibfnamefont {D.~J.}\ \bibnamefont {Dean}},\ }\href@noop
  {} {\enquote {\bibinfo {title} {Coupled-cluster computations of atomic
  nuclei},}\ }\Eprint {http://arxiv.org/abs/arXiv:1312.7872 [nucl-th]}
  {arXiv:1312.7872 [nucl-th]} \BibitemShut {NoStop}%
\bibitem [{\citenamefont {Blaizot}\ and\ \citenamefont
  {Ripka}(1985)}]{BlaizotRipka}%
  \BibitemOpen
  \bibfield  {author} {\bibinfo {author} {\bibfnamefont {J.-P.}\ \bibnamefont
  {Blaizot}}\ and\ \bibinfo {author} {\bibfnamefont {G.}~\bibnamefont
  {Ripka}},\ }\href@noop {} {\emph {\bibinfo {title} {Quantum Theory of Finite
  Systems}}}\ (\bibinfo  {publisher} {The MIT Press},\ \bibinfo {address}
  {Cambridge, MA},\ \bibinfo {year} {1985})\BibitemShut {NoStop}%
\bibitem [{\citenamefont {Ring}\ and\ \citenamefont
  {Schuck}(1980)}]{RingSchuck}%
  \BibitemOpen
  \bibfield  {author} {\bibinfo {author} {\bibfnamefont {P.}~\bibnamefont
  {Ring}}\ and\ \bibinfo {author} {\bibfnamefont {P.}~\bibnamefont {Schuck}},\
  }\href@noop {} {\emph {\bibinfo {title} {The Nuclear Many-Body Problem}}}\
  (\bibinfo  {publisher} {Springer-Verlag},\ \bibinfo {address} {Berlin},\
  \bibinfo {year} {1980})\BibitemShut {NoStop}%
\bibitem [{\citenamefont {Roos}(2005)}]{Roos}%
  \BibitemOpen
  \bibfield  {author} {\bibinfo {author} {\bibfnamefont {B.~O.}\ \bibnamefont
  {Roos}},\ }in\ \href@noop {} {\emph {\bibinfo {booktitle} {Theory and
  Applications of Computational Chemistry: The First Forty Years}}},\ \bibinfo
  {editor} {edited by\ \bibinfo {editor} {\bibfnamefont {C.~E.}\ \bibnamefont
  {Dykstra}}, \bibinfo {editor} {\bibfnamefont {G.}~\bibnamefont {Frenking}},
  \bibinfo {editor} {\bibfnamefont {K.~S.}\ \bibnamefont {Kim}}, \ and\
  \bibinfo {editor} {\bibfnamefont {G.~E.}\ \bibnamefont {Scuseria}}}\
  (\bibinfo  {publisher} {Elsevier},\ \bibinfo {address} {Amsterdam, The
  Netherlands},\ \bibinfo {year} {2005})\ p.\ \bibinfo {pages}
  {725}\BibitemShut {NoStop}%
\bibitem [{\citenamefont {Volya}\ \emph {et~al.}(2001)\citenamefont {Volya},
  \citenamefont {Brown},\ and\ \citenamefont {Zelevinksy}}]{Volya}%
  \BibitemOpen
  \bibfield  {author} {\bibinfo {author} {\bibfnamefont {A.}~\bibnamefont
  {Volya}}, \bibinfo {author} {\bibfnamefont {B.~A.}\ \bibnamefont {Brown}}, \
  and\ \bibinfo {author} {\bibfnamefont {V.}~\bibnamefont {Zelevinksy}},\
  }\href@noop {} {\bibfield  {journal} {\bibinfo  {journal} {Phys. Lett. B}\
  }\textbf {\bibinfo {volume} {509}},\ \bibinfo {pages} {37} (\bibinfo {year}
  {2001})}\BibitemShut {NoStop}%
\bibitem [{\citenamefont {Johnson}\ \emph {et~al.}(2013)\citenamefont
  {Johnson}, \citenamefont {Ayers}, \citenamefont {Limacher}, \citenamefont
  {{De Baerdemacker}}, \citenamefont {{Van Neck}},\ and\ \citenamefont
  {Bultinck}}]{Ayers1}%
  \BibitemOpen
  \bibfield  {author} {\bibinfo {author} {\bibfnamefont {P.~A.}\ \bibnamefont
  {Johnson}}, \bibinfo {author} {\bibfnamefont {P.~W.}\ \bibnamefont {Ayers}},
  \bibinfo {author} {\bibfnamefont {P.~A.}\ \bibnamefont {Limacher}}, \bibinfo
  {author} {\bibfnamefont {S.}~\bibnamefont {{De Baerdemacker}}}, \bibinfo
  {author} {\bibfnamefont {D.}~\bibnamefont {{Van Neck}}}, \ and\ \bibinfo
  {author} {\bibfnamefont {P.}~\bibnamefont {Bultinck}},\ }\href@noop {}
  {\bibfield  {journal} {\bibinfo  {journal} {Comp. Theor. Chem.}\ }\textbf
  {\bibinfo {volume} {1003}},\ \bibinfo {pages} {101} (\bibinfo {year}
  {2013})}\BibitemShut {NoStop}%
\bibitem [{\citenamefont {Limacher}\ \emph {et~al.}(2013)\citenamefont
  {Limacher}, \citenamefont {Ayers}, \citenamefont {Johnson}, \citenamefont
  {{De Baerdemacker}}, \citenamefont {{Van Neck}},\ and\ \citenamefont
  {Bultinck}}]{Ayers2}%
  \BibitemOpen
  \bibfield  {author} {\bibinfo {author} {\bibfnamefont {P.~A.}\ \bibnamefont
  {Limacher}}, \bibinfo {author} {\bibfnamefont {P.~W.}\ \bibnamefont {Ayers}},
  \bibinfo {author} {\bibfnamefont {P.~A.}\ \bibnamefont {Johnson}}, \bibinfo
  {author} {\bibfnamefont {S.}~\bibnamefont {{De Baerdemacker}}}, \bibinfo
  {author} {\bibfnamefont {D.}~\bibnamefont {{Van Neck}}}, \ and\ \bibinfo
  {author} {\bibfnamefont {P.}~\bibnamefont {Bultinck}},\ }\href@noop {}
  {\bibfield  {journal} {\bibinfo  {journal} {J. Chem. Theory Comput.}\
  }\textbf {\bibinfo {volume} {9}},\ \bibinfo {pages} {1394} (\bibinfo {year}
  {2013})}\BibitemShut {NoStop}%
\bibitem [{\citenamefont {Dukelsky}\ \emph {et~al.}(2003)\citenamefont
  {Dukelsky}, \citenamefont {Dussel}, \citenamefont {Hirsch},\ and\
  \citenamefont {Schuck}}]{JorgeBreakdown}%
  \BibitemOpen
  \bibfield  {author} {\bibinfo {author} {\bibfnamefont {J.}~\bibnamefont
  {Dukelsky}}, \bibinfo {author} {\bibfnamefont {G.~G.}\ \bibnamefont
  {Dussel}}, \bibinfo {author} {\bibfnamefont {J.~G.}\ \bibnamefont {Hirsch}},
  \ and\ \bibinfo {author} {\bibfnamefont {P.}~\bibnamefont {Schuck}},\
  }\href@noop {} {\bibfield  {journal} {\bibinfo  {journal} {Nucl. Phys. A}\
  }\textbf {\bibinfo {volume} {714}},\ \bibinfo {pages} {63} (\bibinfo {year}
  {2003})}\BibitemShut {NoStop}%
\bibitem [{\citenamefont {Sandulescu}\ and\ \citenamefont
  {Bertsch}(2008)}]{Sandulescu}%
  \BibitemOpen
  \bibfield  {author} {\bibinfo {author} {\bibfnamefont {N.}~\bibnamefont
  {Sandulescu}}\ and\ \bibinfo {author} {\bibfnamefont {G.~F.}\ \bibnamefont
  {Bertsch}},\ }\href@noop {} {\bibfield  {journal} {\bibinfo  {journal} {Phys.
  Rev. C}\ }\textbf {\bibinfo {volume} {78}},\ \bibinfo {pages} {064318}
  (\bibinfo {year} {2008})}\BibitemShut {NoStop}%
\bibitem [{\citenamefont {Bardeen}\ \emph {et~al.}(1957)\citenamefont
  {Bardeen}, \citenamefont {Cooper},\ and\ \citenamefont
  {Scrieffer}}]{Bardeen}%
  \BibitemOpen
  \bibfield  {author} {\bibinfo {author} {\bibfnamefont {J.}~\bibnamefont
  {Bardeen}}, \bibinfo {author} {\bibfnamefont {L.~N.}\ \bibnamefont {Cooper}},
  \ and\ \bibinfo {author} {\bibfnamefont {J.~R.}\ \bibnamefont {Scrieffer}},\
  }\href@noop {} {\bibfield  {journal} {\bibinfo  {journal} {Phys. Rev.}\
  }\textbf {\bibinfo {volume} {108}},\ \bibinfo {pages} {1175} (\bibinfo {year}
  {1957})}\BibitemShut {NoStop}%
\bibitem [{\citenamefont {Bohr}\ \emph {et~al.}(1958)\citenamefont {Bohr},
  \citenamefont {Mottelson},\ and\ \citenamefont {Pines}}]{Pines}%
  \BibitemOpen
  \bibfield  {author} {\bibinfo {author} {\bibfnamefont {A.}~\bibnamefont
  {Bohr}}, \bibinfo {author} {\bibfnamefont {B.~R.}\ \bibnamefont {Mottelson}},
  \ and\ \bibinfo {author} {\bibfnamefont {D.}~\bibnamefont {Pines}},\
  }\href@noop {} {\bibfield  {journal} {\bibinfo  {journal} {Phys. Rev.}\
  }\textbf {\bibinfo {volume} {110}},\ \bibinfo {pages} {936} (\bibinfo {year}
  {1958})}\BibitemShut {NoStop}%
\bibitem [{\citenamefont {Dietrich}\ \emph {et~al.}(1964)\citenamefont
  {Dietrich}, \citenamefont {Mang},\ and\ \citenamefont {Pradal}}]{Pradal}%
  \BibitemOpen
  \bibfield  {author} {\bibinfo {author} {\bibfnamefont {K.}~\bibnamefont
  {Dietrich}}, \bibinfo {author} {\bibfnamefont {H.~J.}\ \bibnamefont {Mang}},
  \ and\ \bibinfo {author} {\bibfnamefont {J.~H.}\ \bibnamefont {Pradal}},\
  }\href@noop {} {\bibfield  {journal} {\bibinfo  {journal} {Phys. Rev.}\
  }\textbf {\bibinfo {volume} {135}},\ \bibinfo {pages} {22} (\bibinfo {year}
  {1964})}\BibitemShut {NoStop}%
\bibitem [{\citenamefont {Richardson}(1963)}]{Richardson1}%
  \BibitemOpen
  \bibfield  {author} {\bibinfo {author} {\bibfnamefont {R.~W.}\ \bibnamefont
  {Richardson}},\ }\href@noop {} {\bibfield  {journal} {\bibinfo  {journal}
  {Phys. Lett.}\ }\textbf {\bibinfo {volume} {3}},\ \bibinfo {pages} {277}
  (\bibinfo {year} {1963})}\BibitemShut {NoStop}%
\bibitem [{\citenamefont {Richardson}(1966)}]{Richardson4}%
  \BibitemOpen
  \bibfield  {author} {\bibinfo {author} {\bibfnamefont {R.~W.}\ \bibnamefont
  {Richardson}},\ }\href@noop {} {\bibfield  {journal} {\bibinfo  {journal}
  {Phys. Rev.}\ }\textbf {\bibinfo {volume} {141}},\ \bibinfo {pages} {949}
  (\bibinfo {year} {1966})}\BibitemShut {NoStop}%
\bibitem [{\citenamefont {Sierra}\ \emph {et~al.}(2000)\citenamefont {Sierra},
  \citenamefont {Dukelsky}, \citenamefont {Dussel}, \citenamefont {{von
  Delft}},\ and\ \citenamefont {Braun}}]{Sierra}%
  \BibitemOpen
  \bibfield  {author} {\bibinfo {author} {\bibfnamefont {G.}~\bibnamefont
  {Sierra}}, \bibinfo {author} {\bibfnamefont {J.}~\bibnamefont {Dukelsky}},
  \bibinfo {author} {\bibfnamefont {G.~G.}\ \bibnamefont {Dussel}}, \bibinfo
  {author} {\bibfnamefont {J.}~\bibnamefont {{von Delft}}}, \ and\ \bibinfo
  {author} {\bibfnamefont {F.}~\bibnamefont {Braun}},\ }\href@noop {}
  {\bibfield  {journal} {\bibinfo  {journal} {Phys. Rev. B}\ }\textbf {\bibinfo
  {volume} {61}},\ \bibinfo {pages} {11890} (\bibinfo {year}
  {2000})}\BibitemShut {NoStop}%
\bibitem [{\citenamefont {Braun}\ and\ \citenamefont {{von
  Delft}}(1998)}]{Braun}%
  \BibitemOpen
  \bibfield  {author} {\bibinfo {author} {\bibfnamefont {F.}~\bibnamefont
  {Braun}}\ and\ \bibinfo {author} {\bibfnamefont {J.}~\bibnamefont {{von
  Delft}}},\ }\href@noop {} {\bibfield  {journal} {\bibinfo  {journal} {Phys.
  Rev. Lett.}\ }\textbf {\bibinfo {volume} {81}},\ \bibinfo {pages} {4712}
  (\bibinfo {year} {1998})}\BibitemShut {NoStop}%
\bibitem [{\citenamefont {Dukelsky}\ and\ \citenamefont
  {Sierra}(1999)}]{Dukelsky}%
  \BibitemOpen
  \bibfield  {author} {\bibinfo {author} {\bibfnamefont {J.}~\bibnamefont
  {Dukelsky}}\ and\ \bibinfo {author} {\bibfnamefont {G.}~\bibnamefont
  {Sierra}},\ }\href@noop {} {\bibfield  {journal} {\bibinfo  {journal} {Phys.
  Rev. Lett.}\ }\textbf {\bibinfo {volume} {83}},\ \bibinfo {pages} {172}
  (\bibinfo {year} {1999})}\BibitemShut {NoStop}%
\bibitem [{\citenamefont {Signoracci}\ \emph {et~al.}()\citenamefont
  {Signoracci}, \citenamefont {Duguet},\ and\ \citenamefont {Hagen}}]{Duguet}%
  \BibitemOpen
  \bibfield  {author} {\bibinfo {author} {\bibfnamefont {A.}~\bibnamefont
  {Signoracci}}, \bibinfo {author} {\bibfnamefont {T.}~\bibnamefont {Duguet}},
  \ and\ \bibinfo {author} {\bibfnamefont {G.}~\bibnamefont {Hagen}},\
  }\href@noop {} {\enquote {\bibinfo {title} {Ab initio {B}ogoliubov
  coupled-cluster theory for open-shell nuclei},}\ }\bibinfo {howpublished} {to
  be submitted}\BibitemShut {NoStop}%
\bibitem [{\citenamefont {Lacroix}\ and\ \citenamefont
  {Gambacurta}(2012)}]{LaCroixGambacurta}%
  \BibitemOpen
  \bibfield  {author} {\bibinfo {author} {\bibfnamefont {D.}~\bibnamefont
  {Lacroix}}\ and\ \bibinfo {author} {\bibfnamefont {D.}~\bibnamefont
  {Gambacurta}},\ }\href@noop {} {\bibfield  {journal} {\bibinfo  {journal}
  {Phys. Rev. C}\ }\textbf {\bibinfo {volume} {86}},\ \bibinfo {pages} {014306}
  (\bibinfo {year} {2012})}\BibitemShut {NoStop}%
\bibitem [{\citenamefont {Stolarczyk}\ and\ \citenamefont
  {Monkhorst}(2010)}]{StolarczykMonkhorst}%
  \BibitemOpen
  \bibfield  {author} {\bibinfo {author} {\bibfnamefont {L.}~\bibnamefont
  {Stolarczyk}}\ and\ \bibinfo {author} {\bibfnamefont {H.}~\bibnamefont
  {Monkhorst}},\ }\href@noop {} {\bibfield  {journal} {\bibinfo  {journal}
  {Mol. Phys.}\ }\textbf {\bibinfo {volume} {108}},\ \bibinfo {pages} {3067}
  (\bibinfo {year} {2010})}\BibitemShut {NoStop}%
\bibitem [{\citenamefont {Brueckner}\ and\ \citenamefont
  {Wada}(1956)}]{Brueckner}%
  \BibitemOpen
  \bibfield  {author} {\bibinfo {author} {\bibfnamefont {K.~A.}\ \bibnamefont
  {Brueckner}}\ and\ \bibinfo {author} {\bibfnamefont {W.}~\bibnamefont
  {Wada}},\ }\href@noop {} {\bibfield  {journal} {\bibinfo  {journal} {Phys.
  Rev.}\ }\textbf {\bibinfo {volume} {103}},\ \bibinfo {pages} {1008} (\bibinfo
  {year} {1956})}\BibitemShut {NoStop}%
\bibitem [{\citenamefont {Hirsch}\ \emph {et~al.}(2002)\citenamefont {Hirsch},
  \citenamefont {Mariano}, \citenamefont {Dukelsky},\ and\ \citenamefont
  {Schuck}}]{ParticleHoleSymmetry}%
  \BibitemOpen
  \bibfield  {author} {\bibinfo {author} {\bibfnamefont {J.~G.}\ \bibnamefont
  {Hirsch}}, \bibinfo {author} {\bibfnamefont {A.}~\bibnamefont {Mariano}},
  \bibinfo {author} {\bibfnamefont {J.}~\bibnamefont {Dukelsky}}, \ and\
  \bibinfo {author} {\bibfnamefont {P.}~\bibnamefont {Schuck}},\ }\href@noop {}
  {\bibfield  {journal} {\bibinfo  {journal} {Ann. Phys.}\ }\textbf {\bibinfo
  {volume} {296}},\ \bibinfo {pages} {187} (\bibinfo {year}
  {2002})}\BibitemShut {NoStop}%
\bibitem [{\citenamefont {Staroverov}\ and\ \citenamefont
  {Scuseria}(2002)}]{staroverov2002}%
  \BibitemOpen
  \bibfield  {author} {\bibinfo {author} {\bibfnamefont {V.~N.}\ \bibnamefont
  {Staroverov}}\ and\ \bibinfo {author} {\bibfnamefont {G.~E.}\ \bibnamefont
  {Scuseria}},\ }\href@noop {} {\bibfield  {journal} {\bibinfo  {journal} {J.
  Chem. Phys.}\ }\textbf {\bibinfo {volume} {117}},\ \bibinfo {pages} {11107}
  (\bibinfo {year} {2002})}\BibitemShut {NoStop}%
\bibitem [{\citenamefont {Stanton}\ \emph {et~al.}(1992)\citenamefont
  {Stanton}, \citenamefont {Gauss},\ and\ \citenamefont
  {Bartlett}}]{BruecknerSymm1}%
  \BibitemOpen
  \bibfield  {author} {\bibinfo {author} {\bibfnamefont {J.~F.}\ \bibnamefont
  {Stanton}}, \bibinfo {author} {\bibfnamefont {J.}~\bibnamefont {Gauss}}, \
  and\ \bibinfo {author} {\bibfnamefont {R.~J.}\ \bibnamefont {Bartlett}},\
  }\href@noop {} {\bibfield  {journal} {\bibinfo  {journal} {J. Chem. Phys.}\
  }\textbf {\bibinfo {volume} {97}},\ \bibinfo {pages} {5554} (\bibinfo {year}
  {1992})}\BibitemShut {NoStop}%
\bibitem [{\citenamefont {Barnes}\ and\ \citenamefont
  {Lindh}(1994)}]{BruecknerSymm2}%
  \BibitemOpen
  \bibfield  {author} {\bibinfo {author} {\bibfnamefont {L.~A.}\ \bibnamefont
  {Barnes}}\ and\ \bibinfo {author} {\bibfnamefont {R.}~\bibnamefont {Lindh}},\
  }\href@noop {} {\bibfield  {journal} {\bibinfo  {journal} {Chem. Phys.
  Lett.}\ }\textbf {\bibinfo {volume} {223}},\ \bibinfo {pages} {207} (\bibinfo
  {year} {1994})}\BibitemShut {NoStop}%
\bibitem [{\citenamefont {Xie}\ \emph {et~al.}(1996)\citenamefont {Xie},
  \citenamefont {Aellen}, \citenamefont {Yamaguchi},\ and\ \citenamefont
  {Schaefer}}]{BruecknerSymm3}%
  \BibitemOpen
  \bibfield  {author} {\bibinfo {author} {\bibfnamefont {Y.}~\bibnamefont
  {Xie}}, \bibinfo {author} {\bibfnamefont {W.~D.}\ \bibnamefont {Aellen}},
  \bibinfo {author} {\bibfnamefont {Y.}~\bibnamefont {Yamaguchi}}, \ and\
  \bibinfo {author} {\bibfnamefont {H.~F.}\ \bibnamefont {Schaefer}},\
  }\href@noop {} {\bibfield  {journal} {\bibinfo  {journal} {J. Chem. Phys.}\
  }\textbf {\bibinfo {volume} {104}},\ \bibinfo {pages} {7615} (\bibinfo {year}
  {1996})}\BibitemShut {NoStop}%
\bibitem [{\citenamefont {Crawford}\ \emph {et~al.}(1997)\citenamefont
  {Crawford}, \citenamefont {Stanton}, \citenamefont {Szalay},\ and\
  \citenamefont {Shaefer}}]{BruecknerSymm4}%
  \BibitemOpen
  \bibfield  {author} {\bibinfo {author} {\bibfnamefont {T.~D.}\ \bibnamefont
  {Crawford}}, \bibinfo {author} {\bibfnamefont {J.~F.}\ \bibnamefont
  {Stanton}}, \bibinfo {author} {\bibfnamefont {P.~G.}\ \bibnamefont {Szalay}},
  \ and\ \bibinfo {author} {\bibfnamefont {H.~F.}\ \bibnamefont {Shaefer}},\
  }\href@noop {} {\bibfield  {journal} {\bibinfo  {journal} {J. Chem. Phys.}\
  }\textbf {\bibinfo {volume} {107}},\ \bibinfo {pages} {2525} (\bibinfo {year}
  {1997})}\BibitemShut {NoStop}%
\bibitem [{\citenamefont {{von Delft}}\ \emph {et~al.}(1996)\citenamefont {{von
  Delft}}, \citenamefont {Zaikin}, \citenamefont {Golubev},\ and\ \citenamefont
  {Tichy}}]{Tichy}%
  \BibitemOpen
  \bibfield  {author} {\bibinfo {author} {\bibfnamefont {J.}~\bibnamefont {{von
  Delft}}}, \bibinfo {author} {\bibfnamefont {A.~D.}\ \bibnamefont {Zaikin}},
  \bibinfo {author} {\bibfnamefont {D.~S.}\ \bibnamefont {Golubev}}, \ and\
  \bibinfo {author} {\bibfnamefont {W.}~\bibnamefont {Tichy}},\ }\href@noop {}
  {\bibfield  {journal} {\bibinfo  {journal} {Phys. Rev. Lett.}\ }\textbf
  {\bibinfo {volume} {77}},\ \bibinfo {pages} {3189} (\bibinfo {year}
  {1996})}\BibitemShut {NoStop}%
\bibitem [{\citenamefont {Dukelsky}\ and\ \citenamefont
  {Sierra}(2000)}]{ConstantBCSGap}%
  \BibitemOpen
  \bibfield  {author} {\bibinfo {author} {\bibfnamefont {J.}~\bibnamefont
  {Dukelsky}}\ and\ \bibinfo {author} {\bibfnamefont {G.}~\bibnamefont
  {Sierra}},\ }\href@noop {} {\bibfield  {journal} {\bibinfo  {journal} {Phys.
  Rev. B}\ }\textbf {\bibinfo {volume} {61}},\ \bibinfo {pages} {12302}
  (\bibinfo {year} {2000})}\BibitemShut {NoStop}%
\end{thebibliography}%

\end{document}